\documentclass[%
reprint,
superscriptaddress,
showpacs,
amsmath,amssymb,
aps,
prl,
Ifloatfix,
citeautoscript,
]{revtex4-1}
\usepackage{CJK}
\usepackage[UKenglish,english]{babel}
\usepackage{graphicx}
\usepackage{epstopdf}
\usepackage{mathrsfs}
\usepackage{cleveref}
\usepackage{float}
\usepackage{tensor}
\usepackage{siunitx}
\usepackage{graphicx}
\usepackage{dcolumn}
\usepackage{bm}
\usepackage[normalem]{ulem}
\usepackage{color}
\usepackage[inline]{enumitem}

\providecommand{\diff}[1]{\ensuremath{\mathrm{d}{#1}}}
\providecommand{\rbracket}[1]{\ensuremath{\left( #1 \right)}}
\providecommand{\frbracket}[1]{\ensuremath{\!\left( #1 \right)}}
\providecommand{\sbracket}[1]{\ensuremath{\left[ #1 \right]}}

\providecommand{\cbracket}[1]{\ensuremath{\left\{ #1 \right\}}}

\usepackage{tensor}
\usepackage{braket}

\graphicspath{{../}}
\begin{document}
	\begin{CJK*}{UTF8}{}

		\title{Violation of a causal inequality in a spacetime with definite causal order}
		\CJKfamily{bsmi}
		\author{C. T. Marco Ho \CJKkern(何宗泰)}
		\email{c.ho1@uq.edu.au}
		\affiliation{Centre for Quantum Computation and Communication Technology, School of Mathematics and Physics, University of Queensland, Brisbane, Queensland 4072, Australia}
	
	\author{Fabio Costa}
	\affiliation{Centre for Engineered Quantum Systems, School of Mathematics and Physics,
The University of Queensland, St. Lucia, QLD 4072, Australia}

\author{Christina Giarmatzi}
	\affiliation{Centre for Engineered Quantum Systems, School of Mathematics and Physics,
The University of Queensland, St. Lucia, QLD 4072, Australia}
		\affiliation{Centre for Quantum Computation and Communication Technology, School of Mathematics and Physics, University of Queensland, Brisbane, Queensland 4072, Australia}
		
		\author{Timothy C. Ralph}%
		\email{ralph@physics.uq.edu.au}
		\affiliation{Centre for Quantum Computation and Communication Technology, School of Mathematics and Physics, University of Queensland, Brisbane, Queensland 4072, Australia}
		
		\date{\today}
		
		\begin{abstract}
Processes with an indefinite causal structure may violate a causal inequality, which quantifies quantum correlations that arise from a lack of causal order. In this paper, we show that when the inequalities are analysed with a Gaussian-localised field theoretic definition of particles and labs, the causal indeterminacy of the fields themselves allows a causal inequality to be violated within the causal structure of Minkowski spacetime. We quantify the violation of the inequality and determine the optimal ordering of observers.
			
		\end{abstract}
		\maketitle
	\end{CJK*}


It is customary to think of physical processes and phenomena as built from events with definite causal relations. Recently, there has been great interest in whether more general causal structures are possible. A main motivation is the expectation that a fundamental theory combining the indeterminacy of quantum physics and the dynamical causal structure of general relativity should include \emph{indefinite causal structures}~\cite{hardy_towards_2007, brukner_quantum_2014}. Processes with indefinite causal structure have also been proposed as possible resources for a variety of tasks \cite{chiribella_quantum_2013, chiribella_perfect_2012,araujo_computational_2014, feix_quantum_2015, guerin_exponential_2016, ebler_enhanced_2018}, with an ongoing effort towards their practical realisation~\cite{procopio_experimental_2015, rubino_experimental_2017, rubino_experimental_2017-1, goswami_indefinite_2018}.

The correlations between events in a definite causal structure satisfy  \emph{causal inequalities} \cite{oreshkov_quantum_2012, oreshkov_causal_2016, branciard_simplest_2016, abbott_multipartite_2016}, derived from the assumption that only one-way signalling is possible: if an event $A$ is the cause of an event $B$, then $B$ cannot be the cause of $A$. A violation of such inequalities would imply that no definite causal order between the events exists. It has been shown that it is possible to violate the causal inequalities within a framework that only assumes the local validity of quantum theory but makes no assumptions regarding a possible background causal structure~\cite{oreshkov_quantum_2012}. However, the physical interpretation of the framework is still uncertain.

In practice, a causal inequality could be violated trivially simply by allowing parties to exchange information across an extended period of time; any probability distribution can be obtained in this way. The interest in the subject derives from the possibility that the inequalities might be violated under stricter conditions, thus demonstrating genuinely new types of causal relations. In Ref.~\cite{oreshkov_quantum_2012} these conditions were proposed to be that of
\emph{closed laboratories}---each event is generated through a single operation on a physical system, which cannot interact with the outside world during the operation---and of \emph{free choice}---an experimenter can perform an arbitrary operation in the closed lab and the choice of operation is independent of other variable relevant to the system under investigation. To date no physical process has been proposed that can violate causal inequalities under such conditions.

We propose a protocol where two parties can violate a causal inequality by acting on Gaussian-localised field modes of photons in Minkowski spacetime. This is possible because operations on the modes are extended in time, so that each intersects the future light-cone of the other. Such laboratories which perform the operations are strictly localised in space and their operations are temporally extended and centred around a spacetime event which is used as a label for the operation. For example, a physical lab (henceforth referred to as `laboratory' or `lab') is a space of finite spatial extent $\Delta x_\text{lab}$ much smaller than the distance to other labs that performs operations centred at $(t_\text{lab}, x_\text{lab})$ on certain specified modes. We thus define `laboratory' as the physical space and `closed laboratory' as the physical space and its operations on certain specified modes. However, if we take the space-time location of the operations---rather than their action on modes---as a definition of laboratories, and identify `closed' labs with compact, space-like separated regions, we would conclude that the violation is due to the failure of the closed lab condition. We comment how this latter perspective is problematic, since any finite-energy mode is necessarily temporally extended, and a small violation of the inequalities is always possible. Thus, we argue that from an operational point of view, freely chosen operations on the modes provide a realisation of closed laboratories, satisfying the conditions for a genuine violation of the inequalities.

\textbf{\emph{Causal inequalities}}---We consider two parties, $A$ (Alice) and $B$ (Bob), who receive classical inputs $x$, $y$ and generate classical outputs $a$, $b$, respectively. For simplicity, we restrict to binary variables and assume that the inputs are uniformly distributed, $P(x,y)=\frac{1}{4}$ for any pair of values $x$, $y$.

The goal for the parties is to guess each other's input, i.e., to maximise the probability \cite{branciard_simplest_2016}
\begin{equation}
P_\text{succ} = \frac{1}{2} \sbracket{P\rbracket{x=b}+P\rbracket{y=a}}.
\label{inequality}
\end{equation}

A definite causal order between the labs imposes constraints on the probability of success: if Alice can signal to Bob, Bob cannot signal to Alice and vice versa. Even if the causal order between the labs is unknown, or decided with some probability by some external variables, the probability of success is bounded by the \emph{causal inequality}~\cite{branciard_simplest_2016}
\begin{equation}
P_\text{succ}\leq \frac{3}{4}.
\label{max}
\end{equation}

This inequality (a simplified version of the original \cite{oreshkov_quantum_2012}) must be satisfied if the operations producing the correlations are each performed between two time instants, defined with respect to a background causal structure, and the system on which Alice (Bob) performs the operation is isolated from the outside world between those two instants.
In a quantum setting, the times at which operations are performed can be subject to indeterminacy. This opens the possibility of violating a causal inequality with operations that still satisfy a reasonable `closed laboratory' assumption. 
As sketched in Ref.~\cite{oreshkov_quantum_2012}, a `closed lab' can be operationally defined---without reference to a background causal structure---in terms of the possible operations that can be performed in it. If a party is free to choose any operation that formally transforms an input Hilbert space to an output Hilbert space, and each operation can in principle be verified through tomography by external parties feeding appropriate states and performing appropriate measurements, we say---by definition---that the party acts in a closed lab. Crucially, the input and output Hilbert spaces do not have to be identified with instants in time: even when a background spacetime structure is assumed, quantum labs can be delocalised in time~\cite{oreshkov_whereabouts_2018}.

\textbf{\emph{Violation of causal inequality with field modes}}---We now present a scenario that, by exploiting temporally-delocalised field modes, enables the violation of the above inequality while satisfying the closed-laboratory assumption. In particular, we consider Gaussian-localised single-particle excitations of optical field modes in Minkowski spacetime,
\begin{equation}
\Ket{1,j} = a_j^\dagger (t,x)\Ket{0}, \label{eq:singleparticle}
\end{equation}
where $j = h, v$ is a polarisation index and the mode is defined by a Gaussian superposition of plane wave modes with annihilation operators
\begin{equation}
a_j(t,x) = \int \diff{k} \frac{e^{-\frac{\rbracket{k-k_0}^2}{4\sigma^2}}}{\rbracket{2\pi 	\sigma^2}^{\frac{1}{4}}} e^{-i k \rbracket{t - x}} a_{k,j}, 
\label{mode}
\end{equation}
where we use units for which $c = \hbar = 1$, $a_{k,j}$ are single frequency Minkowski operators and $\Ket{0}$ is the Minkowski vacuum which is annihilated ($a_k\Ket{0} = 0,~\forall k$) by the Minkowski operators. Note that Eq.~\eqref{eq:singleparticle} is a pure state and so contains all information about the particle. This Gaussian-localised particle has a central wave number of $k_0$ and is peaked along the trajectory
$k_0(t - x) = 0$ with a spatio-temporal width of $1/\sigma$. More realistically we can also require a transverse Gaussian profile for the mode that localises the particle in the transverse directions as well. However, provided we assume that all operations are carried out close to the focus of the mode then the paraxial approximation implies that the $1 + 1$ dimensional description of the mode in Eq.~\eqref{mode} is a good approximation to the full $3 + 1$ dimensional description.

\begin{figure}
	\centering
	\includegraphics[width=0.4\textwidth]{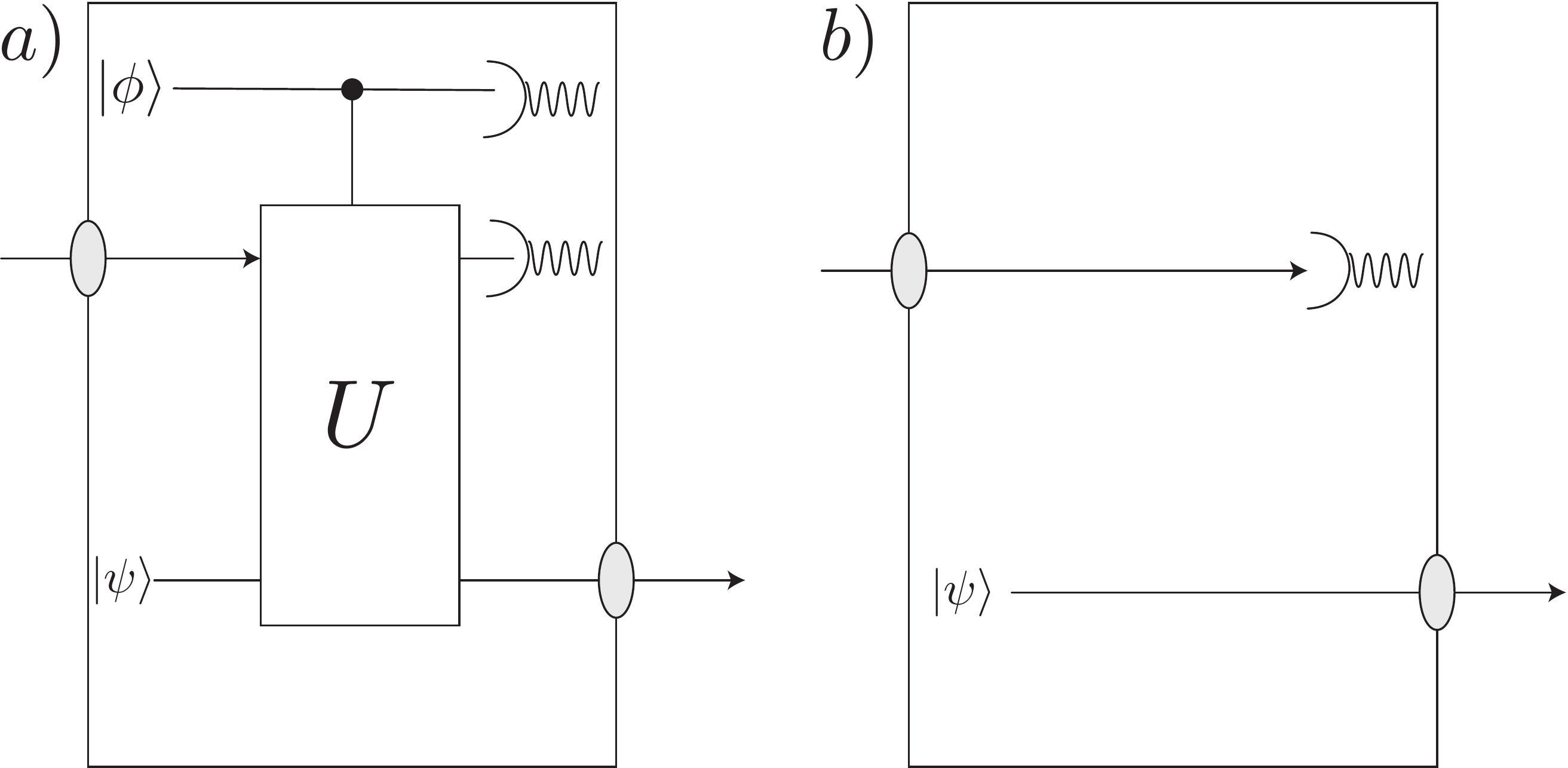}
	\caption{(Color online) a) An illustration possible operations in a lab.  b) The setup we use to violate the inequality has no control qubits and no interactions between input and output.}
	\label{fig:setup}
\end{figure}

A party $A$ (respectively, $B$) that can perform arbitrary operations on---and only on---the single-particle states of such a mode effectively defines a `closed lab'.
To make this definition operationally meaningful, we assume that mode selective mirrors \cite{eckstein_quantum_2011} at the input $I_A$ ($I_B$) and output $O_A$ ($O_B$) allow only a single mode, $\hat a_A$ ($\hat a_B$), to enter and leave Alice's (Bob's) lab (see Fig.\ref{fig:setup}). Note that the labs are finite in spatial extent with a size much smaller than the distance to each other, so the two do not intersect. Modes that are orthogonal to $\hat a_A$ ($\hat a_B$) are completely reflected. In this way the operations in each lab are restricted to a single mode. The operations that act on the mode are centred around an event $(t_X, x_X)$, $X=A, B$. (We assume the mirrors are polarisation insensitive.) Passive mirrors and lenses external to the labs are allowed to direct and focus fields into or away from the labs. 

The closed-lab assumption requires that each party can perform arbitrary operations on the respective single-particle space. Possible operations include unitaries, projective measurements of states $a_j^\dagger(t_\text{X},x_\text{X}) \Ket{0}$, and preparation of states in the same modes. More general operations could require interactions with a local ancilla, e.g., applying a controlled unitary on input and output system followed by a detection of control and input system, Fig.~\ref{fig:setup} a). Interactions with an ancilla do not violate the closed-lab assumption as long as the ancilla is not correlated with any other system outside the lab. Crucially, the assumption can be verified operationally, separately for each lab, by an external party sending selected states to the input mirror and performing measurements at the output. The verifier would then be able to tomographically reconstruct the operations, certifying that each party is indeed free to perform an arbitrary operation on the respective mode. More discussion of the particle states, mode-selective mirrors and lab operations used to define the closed labs can be found in the Supplemental Material at [URL will be inserted by publisher]].

We now consider the specific setup of Fig.~\ref{fig:setup}(b) and assume that Alice and Bob's modes have the same width $\sigma$. In general, this need not be the case, but as we are trying to maximise the violation, this is the simplest choice. Also, for simplicity we assume all operations and detections have unit efficiency. 

\begin{figure}
	\centering
	\includegraphics[width=0.4\textwidth]{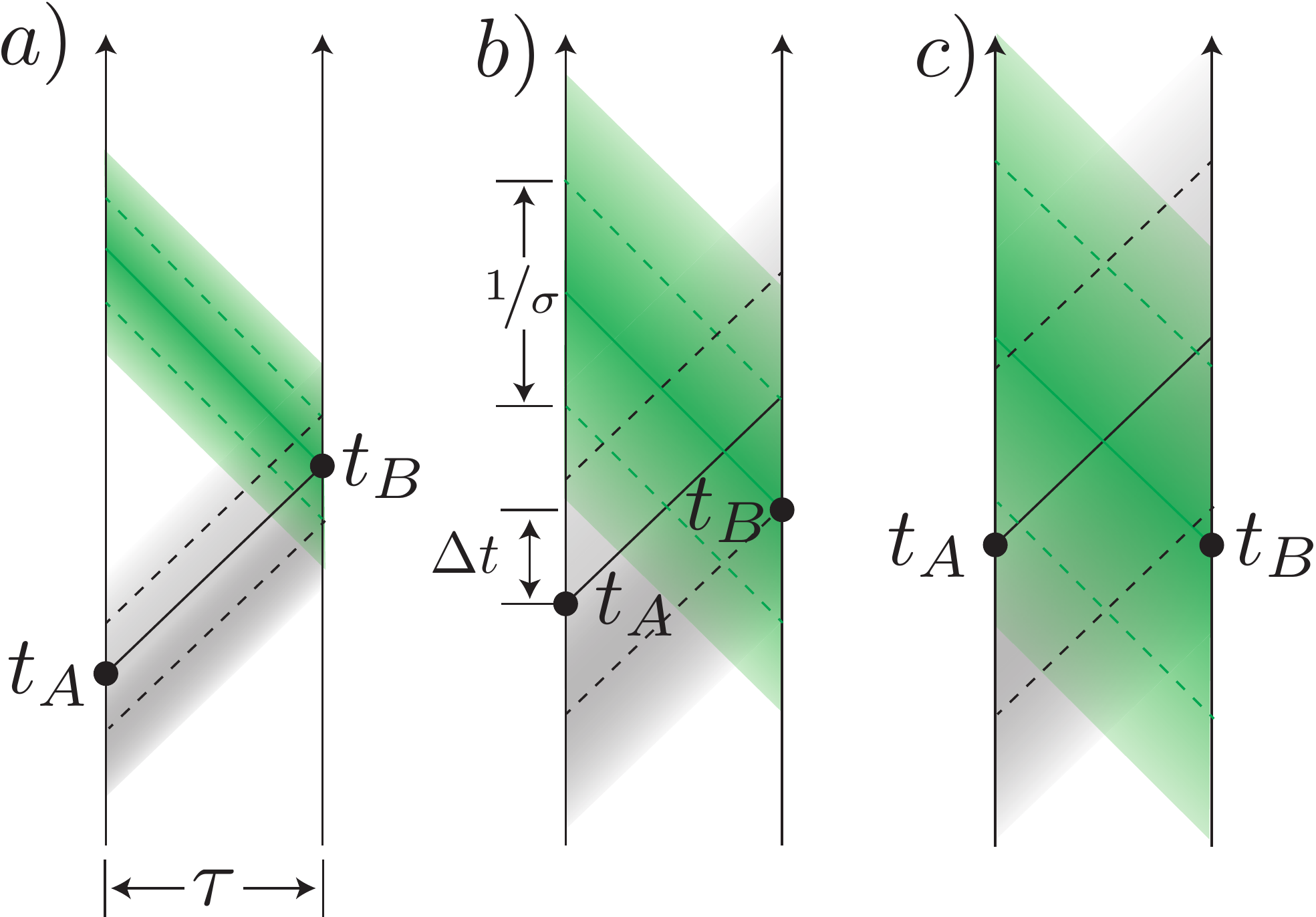}
	\caption{(Color online) A sketch of three regimes. The vertical black lines represent the labs with negligible size on the scale of the diagram. The black dots represent the times $t_A$ and $t_B$. The dotted lines and shaded areas represents the temporal width of the wavepacket $\frac{1}{\sigma}$.  a) is the optimal case when $\sigma \gtrsim \frac{1}{\tau}$ and $\Delta t = \pm\tau$ b) is the optimal case when $\frac{1}{\sqrt{2} \tau}<\sigma \lesssim \frac{1}{\tau}$ and c) is the optimal case when $\sigma \leq \frac{1}{\sqrt{2} \tau}$ and $\Delta t = 0$.}
	\label{modesoverlap}
\end{figure}

The protocol proceeds in the following way. Alice measures the polarisation state of her incoming mode in the horizontal/vertical basis and records her guess $a$ for Bob's bit. Three results are possible: (i) a $h$-polarized photon is detected; (ii) a $v$-polarized photon is detected; (iii) no photon is detected. In case (i) Alice records $a=0$, in case (ii) she records $a=1$, and in case (iii) she randomly chooses to record a zero or one. Simultaneously, Alice prepares the single-photon state: $a_j^{\dagger}(t_A,x_A) |0 \rangle$, choosing the polarisation to be $j = h$ or $v$ according to the value $x=0$ or $x=1$ of the random bit she is trying to send Bob. As the mode of the photon matches the acceptance mode of the output mirror, it escapes from Alice's lab with no attenuation. Bob's protocol is identical except that he measures and prepares the single-photon states $a_j^{\dagger}(t_B,x_B) |0 \rangle$, matching the acceptance mode of his input and output mirrors respectively. We have defined Alice and Bob's modes as right moving modes, i.e., localised on the trajectories $t_X - x_X$. We assume Bob is to the right of Alice (see Fig.~\ref{modesoverlap}) and allow a passive mirror outside Bob's station to reflect Bob's output from right-moving to left-moving. A similar mirror outside Alice's lab reflects left-moving modes back into right moving modes that impinge on Alice's mode selective input mirror. In the following we will ignore the slight asymmetry of this situation and assume the effective propagation distance between the labs is simply $|x_A - x_B|$.

Given our assumptions about the ideal operation of the components it is clear that if Alice (Bob) detects a photon in their polarisation detector they will successfully determine the bit value sent by Bob (Alice). Hence, in order to calculate the value of $P_\text{succ,local}$ (Eq.~\ref{inequality}) we need to determine the probability for Alice (Bob) to detect the photon prepared by Bob (Alice). We can calculate the transmission probability for an excitation of Alice's mode to get through Bob's input mirror via the absolute square of the overlap between their modes:
\begin{align}
P_\text{Bob's mirror} &=\left| \Braket{0|a(t_B,x_B) a^\dagger(t_A,x_A)|0}\right|^2 \\
&= \left|\int \mathrm{d}k ~\frac{e^{-\frac{\rbracket{k-k_0}^2}{2\sigma^2}}}{\rbracket{2\pi 	\sigma^2}^{\frac{1}{2}}} e^{i\sbracket{\omega_k (t_A-t_B) -k\rbracket{x_A-x_B}}}\right|^2\\
& = e^{-\rbracket{t_A-t_B + \tau}^2 \sigma^2}
\end{align}
where $\tau \equiv x_B-x_A$, with the assumption that $k_0\gg \sigma$ and using the usual commutation rule $\sbracket{a_k,a_{k'}^\dagger} = \delta\frbracket{k-k'}$. The above analysis can be repeated for a photon from Bob to Alice and we obtain the similar result:
\begin{align}
P_\text{Alice's mirror} &=\left| \Braket{0|a(t_A,x_A) a^\dagger(t_B,x_B)|0}\right|^2 \\
&= e^{-\rbracket{t_B-t_A + \tau}^2 \sigma^2}.
\end{align}
We can now specify the probability that Bob measures Alice's bit correctly as the probability that the photon is transmitted through Bob's mirror, after which he can definitely know the bit value, plus the probability that the photon is reflected multiplied by the probability he correctly guesses Alice's bit, i.e. $\frac{1}{2}$. Hence we obtain,
\begin{equation}
P(y = a) = e^{-\rbracket{t_A-t_B + \tau}^2 \sigma^2} + \frac{1}{2}\rbracket{1-e^{-\rbracket{t_A-t_B + \tau}^2 \sigma^2}}.
\end{equation}
Similarly for Alice measuring Bob's qubit,
\begin{equation}
P(x= b) = e^{-\rbracket{t_B-t_A + \tau}^2 \sigma^2} + \frac{1}{2}\rbracket{1-e^{-\rbracket{t_B-t_A + \tau}^2 \sigma^2}}.
\end{equation}
The probability of success is therefore,
\begin{equation}
P_\text{succ} = \frac{1}{4} \rbracket{2 +  e^{-\rbracket{t_A-t_B + \tau}^2 \sigma^2} +e^{-\rbracket{t_B-t_A + \tau}^2 \sigma^2} }.
\end{equation}
This is our main result---for any choice of a finite $\sigma$ and $\tau$, timings can be found for which $P_\text{succ}>\frac{3}{4}$ (Eq.~\ref{max}).

We now investigate the optimal $\Delta t \equiv t_B - t_A$ that maximises this probability of success. From the perspective of perfectly localised particles 
this should be the case when $\Delta t = \pm \tau$ but here there is the competing effect of delocalisation. As a result, the best-case scenario depends on the parameters. For $\sigma \gtrsim \frac{1}{\tau}$, it is optimised by $\Delta t \approx \pm \tau$. 
When $\frac{1}{\sqrt{2} \tau}<\sigma \lesssim \frac{1}{\tau}$, the optimal $\Delta t$ is $0< |\Delta t|<\tau$. In this regime, the average send times of Alice and Bob are no longer light-like separated, instead $t_A$ and $t_B$ become increasing more symmetric as $\sigma$ gets smaller.  When $\sigma \leq \frac{1}{\sqrt{2} \tau}$, the optimum separation in time is $\Delta t=0$ where $t_A=t_B$ and we have the symmetric case.
\begin{figure}
	\centering
	\includegraphics[width=0.44\textwidth]{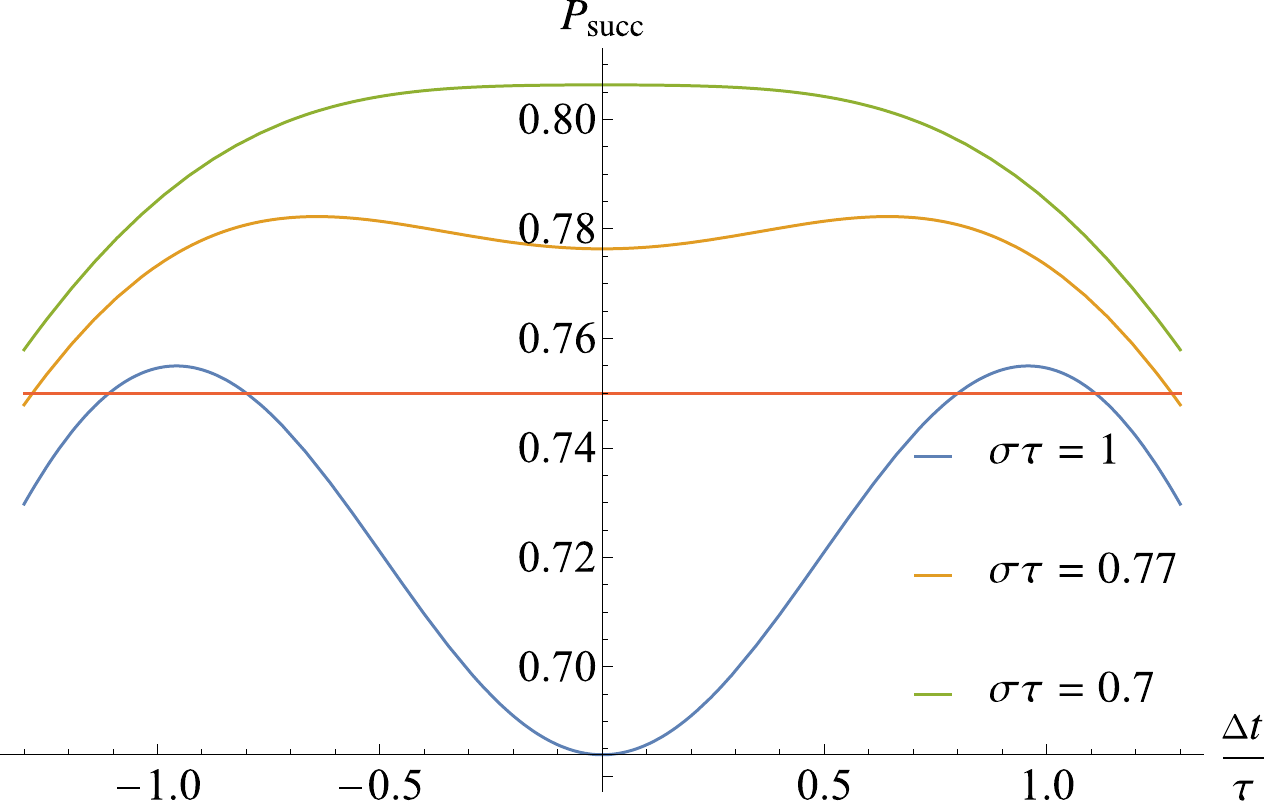}
	\caption{(Color online) The probability of success for three values of $\sigma$ are plotted showing the three regimes where the probability of success is maximised. The red line indicates a probability of success of $0.75$ which can be exceeded for certain choices of $\frac{\Delta t}{\tau}$. As we can see the $\Delta t$ for which the probability is maximised gets smaller as $\sigma \tau$ gets smaller. }
\end{figure}
In the asymmetric case where $\Delta t = \pm\tau$, 
\begin{equation}
P_\text{succ} = \frac{1}{4} \rbracket{3 +  e^{-\rbracket{2 \tau}^2 \sigma^2} }
\end{equation}
and we have a violation of the inequality for any $\sigma < \infty$. In the symmetric case, $\Delta t = 0$ and the probability of success is,
\begin{equation}
P_\text{succ} = \frac{1}{2} \rbracket{1 + e^{-\tau^2 \sigma^2} }
\end{equation}
for which $P_\text{succ}\ge \frac{3}{4}$ when $\sigma \leq \frac{\sqrt{\text{ln} 2}}{\tau}$. In all cases, it is always possible for $P_\text{succ} > \frac{3}{4}$. In the limit of strong photon and lab delocalisation $\sigma\rightarrow 0$, $P_\text{succ}\rightarrow 1$, approaching a maximal violation of the inequality. In the limit $\sigma \to \infty$, we obtain perfect localisation, and we get back the causal inequality where $P_\text{succ} \leq \frac{3}{4}$. However, this is an unphysical limit. In order for our solutions to be valid we require $\sigma \ll k_0$ (this ensures that the mode function doesn't bifurcate into both right {\itshape and} left moving components). As a result $\sigma \to \infty$ implies $k_0 \to \infty$ and hence infinite energy.

\textbf{\emph{Conclusion}}---Causal inequalities represent interesting constraints only if additional conditions are imposed on how the correlations are generated---with no restrictions, it is always possible to  generate arbitrary correlations, without the need of quantum effects or exotic spacetime geometry. Although the inequalities are device and theory independent, the conditions on the protocols are model-dependent and have to rely on additional assumptions. 

Crucial to the original formulation of Ref.~\cite{oreshkov_quantum_2012} is the assumption of closed laboratories, which prevents exploiting simple multi-round protocols. We have considered a possible natural background-independent formalisation of this assumption, namely the identification of closed laboratories with field modes. We have presented a protocol where operations matched to particular field modes enable a violation of a causal inequality.

However, when analysed from the perspective of a background causal structure, the same protocol may seem to violate the closed-laboratory assumption: The two `laboratories' act on delocalised modes and therefore sit in regions that are extended in time, both future and past light cone of each region have a large overlap with the other region, and information can freely travel between the two.

Nonetheless, it is questionable whether it is physically meaningful to take the existence of a background causal structure as a primitive notion. Spacetime points are sometimes a useful abstraction of physical events. In classical physics we often consider (point) particles that are perfectly localised, thus physical events such as `particle enters lab' correspond to a spacetime point/event. Such is not the case for a quantum particle which is always delocalised. Spacetime events are therefore of limited use in quantum physics. Thus, it is perhaps better to consider spacetime events/points as a useful mathematical tool than a primitive constituent of physical theory. With this view, events do not exist on their own: they make sense as relational properties between physical degrees of freedom, quantum fields in our case. It is therefore more meaningful to adopt a background-independent notion of local degrees of freedom. Furthermore, sharply-localised modes are unphysical in quantum field theory as they would be associated with infinite energy~\cite{knight_strict_1961, licht_strict_1963}. Thus, it would never be possible to strictly satisfy the closed lab assumption, as formulated from the background causal structure point of view. This is a manifestation of the well-known problem of localisation in QFT~\cite{hegerfeldt_remark_1974, vazquez_local_2014, schroer_localization_2010} (tightly related with the entanglement in the quantum vacuum~\cite{unruh_notes_1976, summers_vacuum_1985, bombelli_quantum_1986, summers_bells_1987, redhead_more_1995, halvorson_generic_2000, reznik_entanglement_2003, calabrese_entanglement_2004, zych_entanglement_2010, ibnouhsein_renormalized_2014, su_spacetime_2016}), namely the question of which quantum degrees of freedom should be associated with local spacetime regions~\cite{newton_localized_1949, fleming_covariant_1965, segal_anti-locality_1965, fleming_reeh-schlieder_2000, halvorson_reeh-schlieder_2001, piazza_volumes_2007, costa_modeling_2009, cacciatori_renormalized_2009, schroeren_is_2010}. Here we have exposed yet another manifestation of this issue: The localisation problem challenges a meaningful, background-independent definition of causal relations in quantum field theory. A formulation of quantum mechanics with no background causal structure~\cite{oreshkov_quantum_2012} that includes quantum fields will necessarily have to face this issue.

As the violation of a causal inequality is possible with measurements in a fixed basis, the `local operations' cannot be embedded in the `process matrix formalism' in which fixed-basis measurements in a bipartite scenario always lead to definite causal order~\cite{oreshkov_quantum_2012, baumann_appearance_2016}.
This leaves open the question of whether, in order to be compatible with field theory, the process matrix formalism needs to be extended to allow for non-linear probabilities or whether the basic structure and the assumption of closed laboratories need to be reformulated in order to exclude such possibilities.

\begin{acknowledgements}
This research was funded in part by the Australian Research Council Centre of Excellence for Quantum Computation and Communication Technology (Project No. CE110001027). F.C. acknowledges support through an Australian Research Council Discovery Early Career Researcher Award (DE170100712). This publication was made possible through the support of a grant from the John Templeton Foundation. The opinions expressed in this publication are those of the authors and do not necessarily reflect the views of the John Templeton Foundation. We acknowledge the traditional owners of the land on which the University of Queensland is situated, the Turrbal and Jagera people.
\end{acknowledgements}
\bibliography{inequalityviolation}
\clearpage
\onecolumngrid
\begin{center}
	\textbf{\large Supplementary Material}
\end{center}
\section*{Energy of a Gaussian-localised particle}
In our protocol we exchange Gaussian-localised single-particle excitations between the labs. If we use the Hamiltonian operator, we can show that these single-particle excitations have finite energy provided they are not strictly localised. The Hamiltonian is,
\begin{equation}
H = \sum_{j}\int \mathrm{d}k \frac{|k|}{(2\pi)^{1/2}} a_{k,j}^\dagger a_{k,j}.
\end{equation} 
We find that the expectation,
\begin{align}
&\Braket{1,j|H|1,j} = \int \frac{\mathrm{d}k}{2\pi \sigma} |k| e^{-\frac{(k-k_0)^2}{2\sigma^2}}\\
&\stackrel{k_0\gg \sigma}{\approx}\frac{\sigma}{2\pi} \rbracket{\sqrt{2\pi} \frac{k_0}{\sigma}  + e^{-\frac{k_0^2}{\sigma^2}} \frac{2\sigma^2}{k_0^2} + \mathcal{O}\sbracket{\rbracket{\frac{\sigma}{k_0}}^3}}
\end{align} 
is finite in energy for $\sigma <\infty$ (i.e. not strictly localised). 

\section*{The Mode selective mirror}

In the main text we modelled the mode selective mirror as a projective measurement onto the lab mode. Here we present a more detailed model of the mirror. Let us consider Alice's lab. Fig. 1 represents the mode selective mirror. A complete set of orthonormal modes, $\cbracket{a_i}$, impinges from the outside. This basis set is chosen such that Alice's lab mode, $a_0$, is a member of the set (this can always be done \cite{rohde_spectral_2007}. A complementary \footnote{Complementary in the sense that one mode becomes the other if their propagation direction is reflected by $90^\circ$} and orthogonal set of modes, $\cbracket{b_i}$, impinges from the inside. An incoming mode, $c_\text{in}$ from the outside can then be decomposed as
\begin{align}
c_\text{in} = \sqrt{\eta} ~a_0 + \sum_{i \ne 0} f_i ~a_i,
\end{align}
where $\sqrt{\eta} = \sbracket{c_\text{in}, a_0^{\dagger}}$ is given by the overlap of $c_\text{in}$ and $a_0$. Also note that $\eta +\sum_{i \ne 0} |f_i|^2 =1$. Alice's mode selective mirror can then be modelled by the direct product of unitaries 
\begin{align}
U = \prod_{i \ne 0} e^{i \frac{\pi}{2} \rbracket{a_i b_i^\dagger + a_i^\dagger b_i}}, \label{eq:unitary}
\end{align}
which reflects all $a_i$ with $i \ne 0$, but transmit $a_0$.
\begin{figure}[h]
	\centering
	\includegraphics[width=0.35\textwidth]{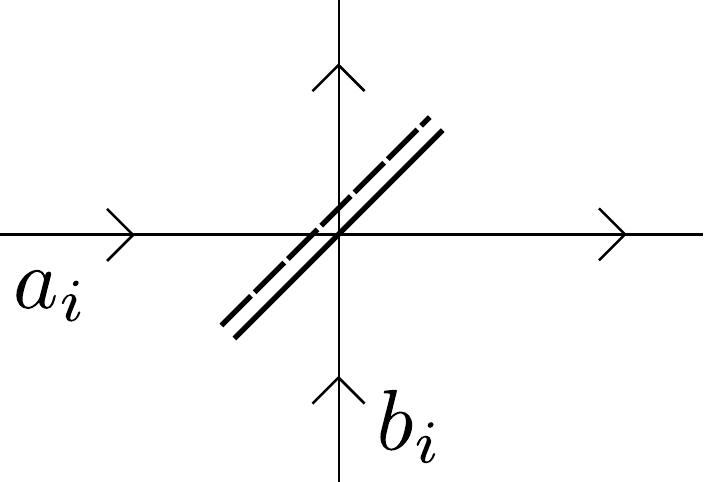}
	\caption{A setup of the mode selective mirror. The beamsplitter is given by the unitary in \cref{eq:unitary}. The state from bob enters from the left and the lab is to the right of the beamsplitter.}
\end{figure}
So a single photon state from Bob, $\rbracket{\Ket{\Psi} =  c_\text{in}^\dagger \Ket{0}}$, going through the mirror becomes
\begin{equation}
U \Ket{\Psi} = U c_\text{in}^\dagger U^{\dagger} U \Ket{0} = \rbracket{\sqrt{\eta}~ a_0^\dagger + i\sum_{i \ne 0} f_i~ b_{i}^\dagger}\Ket{0},
\end{equation}
where we have used that $U^{\dagger} U$ is the identity and $U \Ket{0} = \Ket{0}$. If we trace over the reflected outside modes $b_i$, the reduced density operator of the state in mode $a_0$ inside the lab is,
\begin{equation}
\rho = \eta ~ a_0^\dagger \Ket{0}\Bra{0}  a_0 + (1-\eta) \Ket{0}\Bra{0}.
\end{equation}
All other modes are in the vacuum state. Any operation carried out in the lab will have the maximum probability ($\eta$) of interacting with the photon if it is carried out on the lab mode, $a_0$.
A physical implementation of the mode-selective mirror requires an active interaction such as the pulse gate introduced by \textcite{eckstein_quantum_2011}. 

\section*{Measurements with different timing precision than the mode}

Let us suppose that Alice sends out a mode with a width $\sigma_A$ and Bob tries to measure a mode with a width $\sigma_B$, then we find that
\begin{align}
P_\text{Bob's mirror} &=\left| \Braket{0|a(t_B,x_B,\sigma_B) a^\dagger(t_A,x_A,\sigma_A)|0}\right|^2 \\
& = \frac{2\sigma_A \sigma_B e^{ -\frac{2 (\Delta t +\tau)^2 \sigma_A^2 \sigma_B^2}{\sigma_A^2 + \sigma_B^2}}}{\sigma_A^2 + \sigma_B^2}.
\end{align}
In the case of maximum probability, this gives
\begin{align}
P_\text{Bob's mirror, max} & = \frac{2\sigma_A \sigma_B }{\sigma_A^2 + \sigma_B^2}\\
&=\frac{2 \frac{\sigma_A}{\sigma_B}}{1+ \frac{\sigma_A^2}{\sigma_B^2}}.
\end{align}
Which is strictly $<1$ for $\frac{\sigma_A}{\sigma_B} \neq 1$. The generalised probability of success is therefore
	\begin{equation}
	P_\text{succ} = \frac{1}{4} \rbracket{2 +  \frac{2\sigma_A \sigma_B e^{ -\frac{2 (\Delta t +\tau)^2 \sigma_A^2 \sigma_B^2}{\sigma_A^2 + \sigma_B^2}}}{\sigma_A^2 + \sigma_B^2}+\frac{2\sigma_A \sigma_B e^{ -\frac{2 (-\Delta t +\tau)^2 \sigma_A^2 \sigma_B^2}{\sigma_A^2 + \sigma_B^2}}}{\sigma_A^2 + \sigma_B^2} }.
	\end{equation}
So we see that anything other than $\sigma_A = \sigma_B$ would cause a decrease in the violation of the causal inequality. In particular, the violation would be reduced if Bob tries to measure a mode with greater timing precision (i.e. $\sigma_B>\sigma_A$) than the mode that Alice actually sent. 

\section{Operations in the lab}
We allow all physical operations to be carried out in the labs. However, as previously noted, efficient coupling to any incoming photon will only occur by addressing the lab mode. Similarly, efficient coupling to a photon that will successfully leave the lab via the mode-selective output mirror will only occur by addressing the lab mode. Thus any unitary operations should act specifically on the lab mode or on ancillary states in complementary modes.

There is some subtlety in this, as the physical unitaries doing the operations are localised in space while the mode itself is delocalised. This means that the unitaries are delocalised in time. Such unitaries have a causal order in terms of their central time, or equivalently in terms of their spatial ordering within the lab, but their temporal spread means their operations overlap in time.

However, if all of the above conditions are fulfilled, we could perform any unitary within the lab. This would include measurement and preparing the output state. This is indicated in \cref{fig:setup} a).  In particular, the output state can be prepared conditional on the measurement outcome of the input state, thus justifying the view that---from the laboratory perspective---the measurement causally precedes the preparation. This would not be possible in a protocol where causal inequalities are violated thanks to ``open laboratories'', where a party performs the preparation first and the measurement later, after the system has gone through the other party's lab. 

The violation of the causal inequality indicates that signals can be sent efficiently both from Alice to Bob and from Bob to Alice. As we have commented, preparations of outputs conditional on inputs is allowed by our formalism. One might then worry that this somehow leads to inconsistent behaviour such as Alice sending a message to her own input telling her not to send a message. Of course, our formalism is based on quantum field theory so we expect consistent solutions. The situation we have described is in fact a quantum feedback loop
\cite{yanagisawa_self-consistent_2010}. Whilst in general this problem is very difficult to solve there exists solutions for zero-time feedback loops \cite{combes_slh_2017}. In the next section we investigate a non-trivial loop in the limit of zero-time feedback, where $\tau \sigma \ll 1$ such that the time of travel is much smaller than the temporal spread in the wave packet (i.e. an extreme case of scenario c in the main text).

\section{Feedback loop for a CNOT gate}
\begin{figure}[H]
	\centering
	\includegraphics[width=0.35\columnwidth]{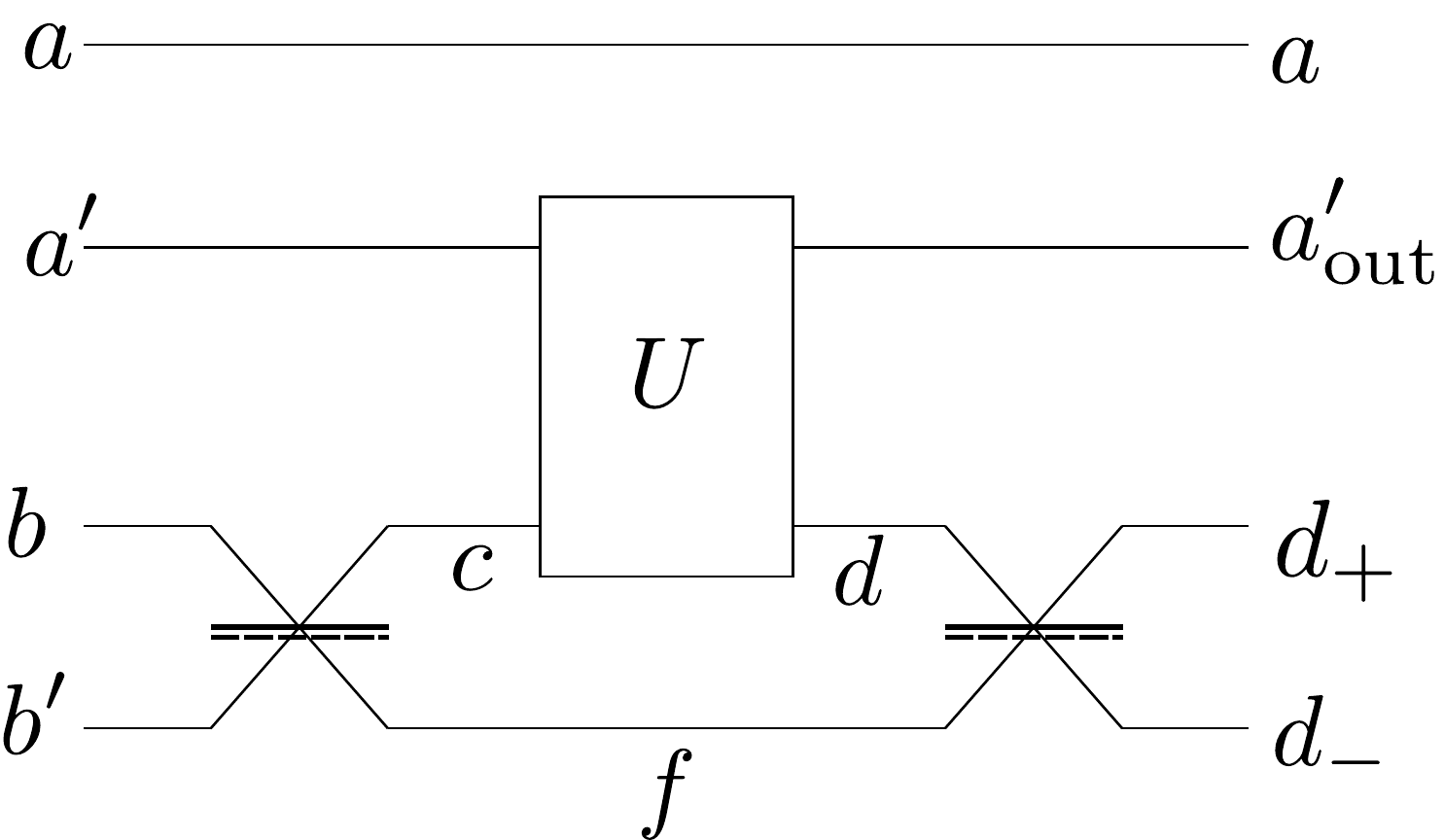}
	\caption{CNOT gate with a cross Kerr non-linearity. The beamsplitters are 50:50. A control qubit is encoded as $\Ket{0_c} = a^\dagger \Ket{0}$ and $\Ket{1_c} = {a'}^\dagger\Ket{0}$. The target qubit is encoded as $\Ket{0_t} = b^\dagger \Ket{0}$ and $\Ket{1_c} = {b'}^\dagger\Ket{0}$.}
	\label{fig:CNOT}
\end{figure}

\begin{figure}[H]
	\centering
	\includegraphics[width=0.35\columnwidth]{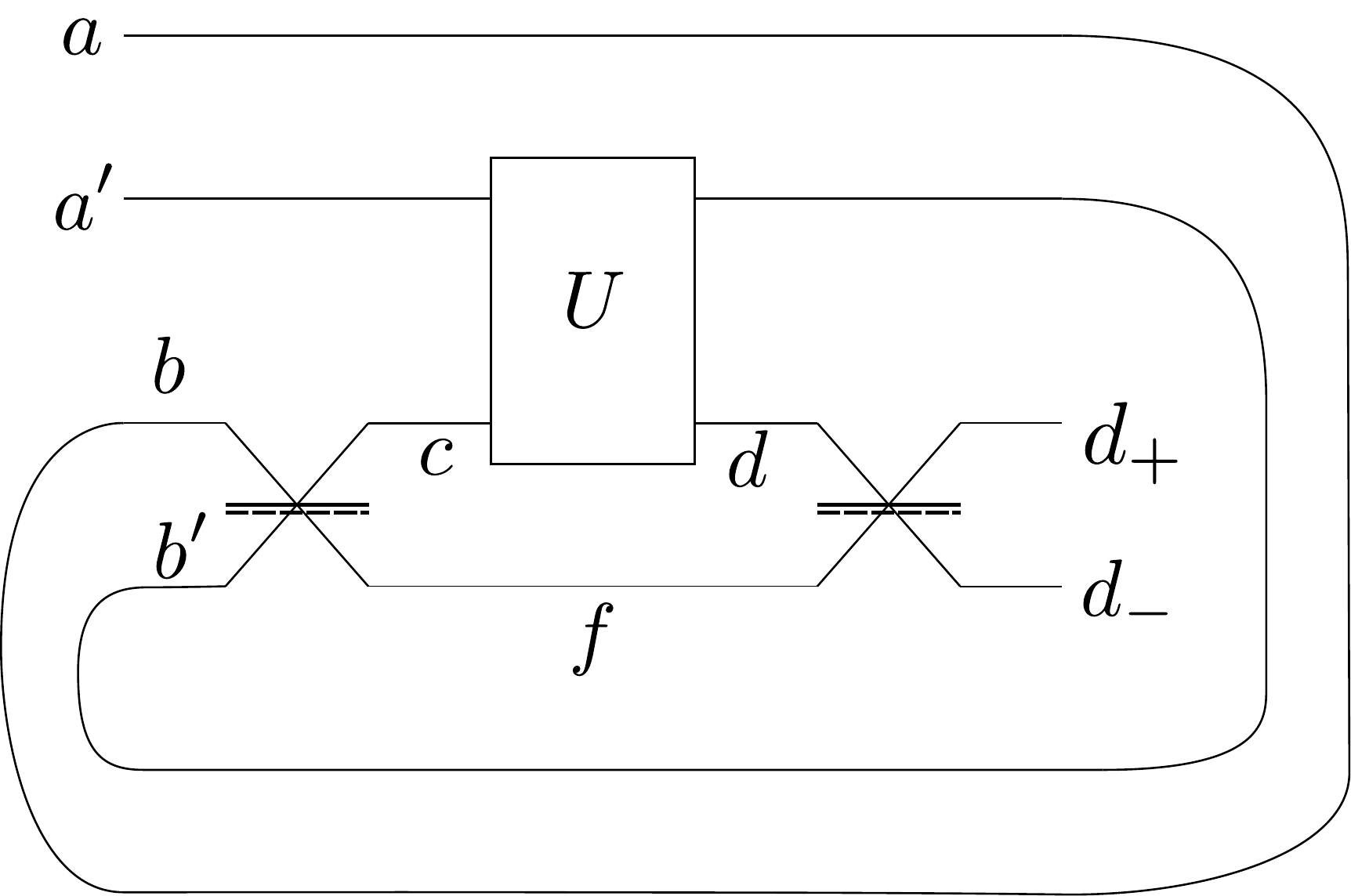}
	\caption{CNOT gate with zero-time feedback.}
	\label{fig:feedback}
\end{figure}

Let us consider a CNOT gate implemented with a cross Kerr non-linearity and dual rail encoding. The CNOT gate is depicted in \cref{fig:CNOT}.
The cross Kerr non-linearity is given by a unitary,
\begin{equation}
U = e^{i \pi c^\dagger c {a'}^\dagger a'}
\end{equation} 
The output of this circuit is,
\begin{align}
a'_\text{out} &= e^{-i \pi c^\dagger c} a'\\
d_\pm &= \frac{1}{2}\sbracket{\rbracket{e^{-i\pi {a'}^\dagger a' } \pm 1 } b + \rbracket{e^{-i\pi {a'}^\dagger a' } \mp 1 } b'}
\end{align}
Now if we feed the output $a$ \& $a'$ to the input $b$ \& $b'$, then we have the circuit in \cref{fig:feedback}. Notice that nominally this assignment can be inconsistent. For example if we prepare the $a$ modes in the state $\Ket{+} = 1/\sqrt{2}\rbracket{\Ket{01}+\Ket{10}}$ and the $b$ modes in the state $\Ket{+}$, then the $a_\text{out}$ modes are in the state $\Ket{-} = 1/\sqrt{2}\rbracket{\Ket{01}-\Ket{10}}$, so the $a_\text{out}$ and $b$ modes seems inconsistent. If we try to fix this by making the $b$ modes in the state $\Ket{-}$ then the $a_\text{out}$ modes switch to $\Ket{+}$ -- seemingly inconsistent again. However, we will see that the actual solution is consistent.
By equating $b = a$ and $b' = a'_\text{out}$ we are assuming the loop is short and the feedback is effectively instantaneous. Notice that we have reduced the Hilbert space of the problem down to 2 dimensions from the previous 4. 
The output is now given by,
\begin{equation}
d_\pm = \frac{1}{2}\sbracket{\rbracket{e^{-i\pi {a'}^\dagger a' } \pm 1 }a  + \rbracket{e^{-i\pi {a'}^\dagger a' } \mp 1 } e^{-i \pi c^\dagger c} a'}
\end{equation}
While we have a self-recursive expression for $c = \frac{1}{\sqrt{2}} \rbracket{a + e^{-i \pi c^\dagger c} a'}$ we will see that we don't need an explicit expression. We can now calculate what this circuit does to logical 0s and 1s.
\begin{align*}
d_\pm a^\dagger \Ket{0} &= \frac{1}{2} \rbracket{1\pm 1} \Ket{0}\\
\implies &\Braket{0| a d_\pm^\dagger d_\pm a^\dagger|0} = \begin{cases}
1 & \text{for} \quad d_+\\
0 & \text{for} \quad d_-
\end{cases}\\
d_\pm {a'}^\dagger \Ket{0} &= \rbracket{e^{-i\pi {a'}^\dagger a' } \mp 1 } e^{-i \pi c^\dagger c} a' {a'}^\dagger \Ket{0}\\
&=\frac{1}{2} \rbracket{1\mp 1} \Ket{0}\\
\implies &\Braket{0| a' d_\pm^\dagger d_\pm {a'}^\dagger|0} = \begin{cases}
0 & \text{for} \quad d_+\\
1 & \text{for} \quad d_-
\end{cases}
\end{align*}
We see that although we do not know the expression for $c$, $e^{-i \pi c^\dagger c}$ acts on the vacuum. For arbitrary input states we find
\begin{align}
&\Braket{0| (\alpha^* a + \beta^* a') d_+^\dagger d_+ (\alpha a^\dagger + \beta a'^\dagger)|0} = |\alpha|^2 \\
&\Braket{0| (\alpha^* a + \beta^* a') d_-^\dagger d_- (\alpha a^\dagger + \beta a'^\dagger)|0} = |\beta|^2
\end{align}
So we see that the zero-time feedback for a CNOT gate (up to a phase rotation) is actually just the identity. 

Let us now consider modes extended in time. For the case in \cref{fig:CNOT}, it is clear how to proceed, we simply specify that the unitary and mirrors are mode matched to modes $a,a', b, b'$. However, when there is a finite-time feedback loop, the modes rentering are shifted in time. If we continue using the mode matched unitary as before, then in the case of scenario a) where the temporal spread of the modes is small compared to the distance between labs, we expect that the unitary would not be matched to the mode by the time most of it propagates back. Therefore in scenario a), we expect that the unitary is also the identity.

\end{document}